# Analysis of the co-design activity: influence of a mixed artifact and contribution of the gestural function in a spatial augmented reality environment

**Maud Poulin**
Univ. Grenoble Alpes, CNRS, Grenoble INP, G-SCOP,
46 Avenue Felix Viallet, 38000 Grenoble, France
maud.poulin@grenoble-inp.fr

**Jean-François Boujut**
Univ. Grenoble Alpes, CNRS, Grenoble INP, G-SCOP,
46 Avenue Felix Viallet, 38000 Grenoble, France
jean-francois.boujut@grenoble-inp.fr

**Cédric Masclet**
Univ. Grenoble Alpes, CNRS, Grenoble INP, G-SCOP,
46 Avenue Felix Viallet, 38000 Grenoble, France
cedric.masclet@gscop.eu

## ABSTRACT

Augmented reality provides new possibilities to propose environments where the designers can take advantage of the physicality of the artifacts while keeping the versatility of digital environments. Mixed objects can therefore provide new media in the interactions between stakeholders. Besides, the increasing interest in user participation in early design phases is limited by the poor representations or the expensive mock ups to be provided in design meetings. Therefore, understanding the role of these mixed artifacts by analyzing and characterizing the interactions is crucial to the development of both design methods and environments. By focusing on multimodal interactions, we aim at providing new results in terms of the design process, in particular by studying the contribution of the gesture in collaborative product co-creativity sessions but also by understanding the role of these multiple interactions in an augmented reality environment.

## KEY-WORDS

Co-design; activity analysis; augmented reality; intermediary objects; multimodal interactions; human system interaction

## INTRODUCTION

The G-SCOP lab and six of its partners were involved in the European project SPARK H2020 (http://spark-project.net/). The goal of the project is to facilitate interactions within co-design sessions involving designers and customers. The project has provided a responsive ICT platform based on Spatial Augmented Reality (SAR) technology. One of the objectives of the SPARK project was to study the influence of a mixed artifact (in a spatial augmented reality environment) on the interactions between designers and clients, and to determine if these influences are beneficial for the overall results from the co-design sessions. This allowed us to focus on gestures made during artifact-centric interactions. Thanks to the development of a real-time quantitative data collection tool and the constitution of an artifact-based interaction coding methodology, we were able to gather information on the type of artefact used by designers and clients during these interactions. Six real co-design sessions were conducted using three different technologies: spatial augmented reality (SAR), augmented reality (AR), and a standard session of tangible artifacts. The results obtained in this project showed that artifact-centric interactions (tangible, numerical, mixed) were more used than unsupported artifact interactions (about 70% of artifact-centric interactions against 30% of ephemeral interactions). Although we have collected results showing a major trend of artifact-centric interactions in contrary to gestures made in the air, we still know little on the use of these latest category. This is why we want to deepen our work on the influence of such a technology, involving a mixed artifact, on the co-design process. For this, we must proceed to an analysis of the co-design activity by the speech and the gesture of which we will strive to define the roles.

## CONTEXT AND PRELIMINARY RESULTS

Co-designing is a large and complex human activity where the problem is still poorly defined, and involves several acceptable solutions at the end of the sessions [1]. Given the co-evolution of problem-solution in design [2], many studies have been conducted to analyze and understand the cognitive activity underlying the design task. Ericsson and Simon [3] are at the origin of the method of analysis of individual protocols whose objective is to understand the cognitive mechanisms and processes that produce relations between the stimulus and the response that appear during human activity. However, verbal interactions remain the most analyzed and traditionally used modality during protocol analysis. Indeed, Jiang and Yen [4] have identified that the use of verbal protocol analysis has significantly increased since the Ericsson and Simon publications and that two types of studies coexist in the literature: the analysis of the individual design and group design analysis with a predominance for the method of "think-aloud" [5]. Wishing to stay closer to the reality of the co-design sessions, think-aloud method does not appear to be relevant in our study because it is more relevant in an individual and experimental design situation.

A study conducted as part of the Eiffel project [6] has provided a method for analyzing the group design process. This method, called COMET, is based on units of sentences where each argument corresponds to a type of spoken action applied to an object to be conceived. The COMET method distinguishes a functional level that examines collaborative design from the point of view of actions and objects implemented in meetings. This method also makes it possible to distinguish a cooperative level showing sequences of actions corresponding to cooperative moves. The application of such a method in our study context seems relevant for its objective is to characterize the structure of functional communications.

Concerning the gestures coding and the functions associated to these gestures, a preliminary work has been carried out in the SPARK project by constituting a gesture coding framework. This coding framework is inspired by the work of McNeill [7] classifying gestures in four different categories: iconic, metaphoric gestures, deictic and beats. In order to quantify and analyze artifact-centric interactions, we also identified deictic gestures (pointing) as relevant. Depending on the type of technology pointed by the participants, these artefact-based pointing gestures were related to the following categories: "tangible" for material artifacts, "digital" for digital artifacts (mainly screens and laptops) and "mixed" for the mixed artifact of the SAR system. Other types of non artefact-centric gestures have also been identified. These are gestures in the air that are not systematically meaningful and that accompany the speech. These gestures, we called ephemeral gestures, were divided into two categories: the gestures of "communication" accompanying simply the speech and the gestures "simulation of an artefact" being mimicry actions of a virtual artefact. These gestures were identified by McNeill as metaphorical and beat gestures (rhythmic) on the one hand, and iconic on the other hand. You will find below a comparative table of these two different coding:

| SPARK coding scheme | McNeill coding scheme |
| --- | --- |
| Virtual artefact | Iconic gestures |
| Communication | Metaphoric gestures |
|  | Beat gestures |
| Tangible/Mixed/Digital | Deictic gestures |
| None | |

Table 1: Comparison of coding scheme classifications between SPARK and McNeill

Thanks to an adaptation of the COMET coding scheme to our study as well as the refinement of the coding of the gestures coming from SPARK, we will study the influence of a mixed artefact, that is to say a spatial augmented reality technology, on the co-design activity.

## PROBLEMATIC AND HYPOTHESES

Based on a rich collection of qualitative information about co-design process in a Spatial Augmented Reality environment, our analysis will focus on the multimodal collaboration.

Then, our main research question focuses on a multimodal analysis method seeking to answer the following research questions:

- "What characterize the tangibility of an artifact on the co-design process ?
- "What does tangibility transform in the collaborative activity ?"

This broad research question includes a subset of more specific questions concerning the specificity of design processes with SAR technology and also, the use and the function of gestures (artefact-centric interactions and ephemeral gestures) in design.

## PROCEDURE

In order to answer the main research question, we will proceed with the following steps.

- In a first part of the thesis, we will analyze the 6 SPARK sessions in a multimodal way to bring out assumptions of tangibility effect of the artefact on the co-design process. For that, we will elaborate the new coding scheme of both speech and gestures. Based on the COMET method for the speech interactions in the one hand and the SPARK coding scheme for the gestures interactions in the other hand, we will work on the combination of these methods to our study. Being interested in the influence of the tangibility of an artifact on a SAR platform, we will compare design sessions on this platform with traditional design sessions. Then, after coding and comparison of the SAR and Standard sessions, activity patterns can be highlighted from both gestural and verbal point of view.

- In a second time, we will create another experimental stage where we will proceed as in the first step of the study but including post-session interviews. These individual post-session interviews based on the explanation interview [8] will allows professionals and researchers to have access to cognitive representation and expressed rationality of the participant through verbalization of his experience and what he perceived he has done during the session. It's not an interpretation of the researcher but what the participant really thought when he made a gesture or when he made an utterance about the artifact during the session. This method uses the work memory and linger on the participants' experience in a defined context, which will be here, the co-design session. The aim of these interviews are to find out what happened (real activity) according to the participant and make implicit knowledge explicit.

    This second experimental stage will give us a new light from the ergonomics and cognitive psychology point of view, complementing the traditional protocol analysis methods, by accessing to the designers and clients' cognitive processes.

    Comparing the design process with a mixed artifact and with a tangible artifact and thanks to the coding of multimodal interactions and the explanation interview, we hope to know if the tangibility of an artifact has an effect on the design activity.

**EXPECTED RESULTS**

By performing a succession of coding and analyses, we seek to understand the designers' activity from a collective standpoint.

Although the SAR and the standard sessions do not use the same technologies, we expect the steps of collaborative design activities [9] to be the same regardless of the tangibility of the artefact. However, some activities would be easier to achieve than others with a SAR technology, for example argue, convince and assess the product being designed.

All these co-design activities use different multimodal interaction categories. For example, we expect that the cognitive synchronisation use more gestures in the air than artefact-centric interactions. Whereas, the argue on a product should use more artefact-centric interactions than gestures in the air. The virtual artifact gestures, as for them, would be very used in cases of justification of ideas by the simulation of an action of a product. Hence, it could highlight privileged associations between activities and nature of artifacts in co-design sessions.

We wish to apply the explanation interview in order to solve the ambiguities of some interactions. Thus, we can adjust our analysis of the co-design process closer to the reality of designers.

**CONCLUSION**

Gestures and speech will be analyzed thanks to two methods: a traditional coding of gestures and speech as done in many protocol analysis researches and the explanation interview from the ergonomic psychology field. We hope to highlight the function of gestures in the co-design cognitive activity and make conclusions about the influence of a mixed artefact on the design process.

**REFERENCES**

1. Simon, H. The structure of ill-defined problems. *Artificial Intelligence*, (1981), 4, 181-201.
2. Dorst, K., & Cross, N. Creativity in the design process: Co-evolution of problem-solution. *Design Studies,* vol 22, 5 (2001), pp :425-437.
3. Ericsson, K. A., & Simon, H. A. Protocol analysis: Verbal reports as data. Cambridge, MA: MIT Press, 1993.
4. Jiang, H. & Yen, C.-C. Protocol analysis in design research: a review. In *Rigor and Relevance in Design*: IASDR 2009, Seoul, Korea October 18_22, Seoul, (2009), pp. 147-156. International Association of Societies of Design Research.
5. Eastman, C.M. Cognitive processes and ill-defined problems: a case study from design. In Proceedings of the First Joint International Conference on Artificial Intelligence. Bedford, MA: MITRE, 1969.
6. Darses, F., Détienne, F., Falzon, P. Visser, W. COMET: A method for Analysing Collective Design Processes (Rapport de Recherche INRIA N° 4258, September 2001).

7. McNeill, D. *Hand and Mind: What gestures reveal about thought*. Chicago: University of Chicago Press, 1992.
8. Vermersch, P. L'entretien d'explicitation. ESF Issy les Moulineaux. Première edition, 1994.
9. Détienne, F., Boujut, J.F., Hohmann, B. Characterization of Collaborative Design and Interaction Management Activities in a Distant Engineering Design Situation. In (Eds) F. Darses, R. Dieng, C. Simone, M. Zacklad, Scenario-based design of collaborative systems, Amsterdam, IOS Press, (2004), pp. 83-98.